%% file: paper.tex
\documentclass[conference,compsoc]{template/IEEEtran}
\IEEEoverridecommandlockouts
\usepackage[T1]{fontenc}
\usepackage{microtype}
\usepackage{amssymb}
\usepackage{cancel}
\usepackage[sort]{cite}
\usepackage[hyphens]{url}
\usepackage{graphicx}
\usepackage[dvipsnames]{xcolor}
\usepackage{balance}

\usepackage[hidelinks]{hyperref}
\usepackage[capitalize,nameinlink,nosort]{cleveref}

\usepackage{xspace}
\usepackage[per-mode=symbol,mode=text,detect-all]{siunitx}
\DeclareSIUnit{\million}{M\@}
\DeclareSIUnit{\thousand}{k\@}
\usepackage[inline]{enumitem}
\usepackage{booktabs}
\usepackage{tabularx}
\usepackage{multirow}

\newif\ifanonymous
\newif\ifauthor
\newif\ifdraft

\anonymoustrue
\anonymousfalse

\authortrue

\drafttrue
\draftfalse

\newcommand{\chk}[1]{{#1}}

\newcommand{\internal}{Self-Contained Data}
\newcommand{\mixed}{Externally-Actionable Data}
\newcommand{\external}{External References}

\newcounter{dilemma}

\newcommand{\dilemmaref}[1]{\underline{\ref{#1}}\xspace}

\newcommand{\ie}{i.e.,\xspace}
\newcommand{\eg}{e.g.,\xspace}

\newcommand{\cf}{cf.\@}

\usepackage{pifont}
\newcommand{\cmark}{\textcolor{ForestGreen}{\ding{51}}}
\newcommand{\xmark}{\textcolor{Red}{\ding{55}}}
\newcommand{\smark}{\textcolor{BurntOrange}{(\ding{51})}}

\newcommand{\impH}{\textcolor{BrickRed}{\textbf{H}}}%
\newcommand{\impM}{\textcolor{YellowOrange}{\textbf{M}}}%
\newcommand{\impL}{\textcolor{OliveGreen}{\textbf{L}}}%

\usepackage{tcolorbox}
\tcbuselibrary{skins,breakable}

\newenvironment{dilemmablock}[2]{%
  \tcolorbox[
    enhanced,
    breakable,
    colback=gray!10,
    colframe=black,
    boxrule=0.5pt,
    left=2mm,right=2mm,top=1mm,bottom=1mm]
  {\bfseries\refstepcounter{dilemma}\dilemmaref{#2}: #1.\label{#2}}}%
  {\endtcolorbox}

    \newcommand{\copyrighttext}{
        \footnotesize
        \textcopyright 2026 IEEE.
        Personal use of this material is permitted.
        Permission from IEEE must be obtained for all other uses, in any current or future media, including reprinting/republishing this material for advertising or promotional purposes, creating new collective works, for resale or redistribution to servers or lists, or reuse of any copyrighted component of this work in other works.
    }
    \newcommand{\copyrightnotice}{%
        \AddToHookNext{shipout/foreground}{%
            \put(0.5\paperwidth,-\paperheight){%
                \makebox(0,0)[b]{\raisebox{25pt}{\parbox[b]{\dimexpr\textwidth-\fboxsep-\fboxrule\relax}{\copyrighttext}}}}}%
    }

\def\BibTeX{{\rm B\kern-.05em{\sc i\kern-.025em b}\kern-.08em
	T\kern-.1667em\lower.7ex\hbox{E}\kern-.125emX}}

\begin{document}

\bstctlcite{IEEEexample:BSTcontrolLong}

\title{Public, Identifiable, and Effectively Immutable:\\ Ethical Lessons from Research on arXiv}

	\author{
	\IEEEauthorblockN{%
		Jan Pennekamp\IEEEauthorrefmark{1}, %
		Johannes Lohmöller\IEEEauthorrefmark{1}, %
		Peter Mayer\IEEEauthorrefmark{2}, %
		Arjun Arunasalam\IEEEauthorrefmark{4} %
	}
	\IEEEauthorblockA{\IEEEauthorrefmark{1}\textit{Communication and Distributed Systems}, RWTH Aachen University, %
	Germany $\cdot$
	\{lastname\}@comsys.rwth-aachen.de}
	\IEEEauthorblockA{\IEEEauthorrefmark{2}\textit{Department of Mathematics and Computer Science}, University of Southern Denmark, Odense, Denmark }%
	\IEEEauthorblockA{\IEEEauthorrefmark{4}\textit{Knight Foundation School of Computing and Information Sciences}, Florida International University, Miami, FL, US }%
	}

\maketitle
\copyrightnotice

\input{paper_content/0_abstract}


\input{paper_content/1_intro}

\input{paper_content/2_background}

\input{paper_content/3_problem}

\input{paper_content/4_assessing_ethics}

\input{paper_content/5_touched_ethical_dilemmas}

\input{paper_content/6_discussion}

\section{Conclusion} %
\label{sec:conclusion}

Public availability alone is not a sufficient basis for ethically unproblematic research use.
In this paper, we examined prior work~\cite{Apruzzeseetal2026X-raying,Pennekampetal2026Hidden,Dubniczkyetal2026You} that utilizes arXiv paper sources as a particularly illustrative case.
Even though the data is intentionally submitted, publicly distributed, directly attributable to named individuals, and effectively immutable, it may still contain sensitive information that was unintentionally disclosed.
This combination creates ethical tensions that are not fully covered by existing frameworks, review processes, or community norms.

Our work stresses that ethical decision-making in cybersecurity research requires more than checking whether data is publicly accessible.
Instead, researchers must treat public availability as the starting point of ethical reasoning, minimize the amplification that their work induces on exposed information, calibrate disclosure to the harm, confidence, and actionability of each finding, and account for the limited remediability that effective immutability entails.
Alongside debating these challenges, we further support future studies by deriving four higher-level dilemmas that warrant consideration.
Collectively, we contribute a resource that aids the design of ethical cybersecurity experimentation.

\section*{Acknowledgment}
\label{sec:acks}

We appreciate the openness of and the valuable exchange with the computer security community when putting together this work:
Fruitful discussions enabled this exhaustive work.
We are further grateful for the valuable feedback from Sophie Stephenson\chk{ and Tadayoshi Kohno}.
We also thank the anonymous reviewers for their valuable feedback and suggestions.

\section*{Research Artifact Statement}
This work did not produce any artifacts.
Its contributions are of an intellectual nature; it discusses works that released their artifacts.
Prior work~\cite{Pennekampetal2026Hidden,Dubniczkyetal2026You} released partial artifacts.

\section*{LLM Usage Statement}

No LLMs or other generative AI tools have been used to prepare this work.
The text has been edited using suggestions provided by Grammarly.
The author(s) manually approved these suggestions and approve this version of the paper.

\balance
\bibliographystyle{IEEEtran}
\bibliography{paper}

\end{document}

%% file: paper_content/0_abstract.tex
\begin{abstract}
Ethical decision-making is central to safe cybersecurity research.
Recent studies, one of which is our own,  analyzing the source files of arXiv papers bring this challenge to light---arXiv publicly distributes the source files, to the surprise of many authors.
Unique aspects %
reveal ethical challenges and dilemmas that extend beyond established ethical frameworks, institutional processes, and community best practices.
In this work, we examine this case %
and provide insights for similar research on  other  public live resources, such as Zenodo or the Wayback Machine.
While such empirical studies are particularly realistic, they must account for  personally identifiable information, leaked secrets, and other sensitive details that could expose, shame, or otherwise negatively affect individuals, possibly leading to both reputational and practical harm.

Drawing on our collective experience and discussions among researchers, affected authors, and readers, we make three contributions:
First, we define the ethical setting and highlight that even intentionally-public data with deliberate real-name attribution can raise ethical concerns when repurposed for empirical analysis.
Second, we discuss the ethical issues surrounding the use of arXiv data and derive four higher-level dilemmas that researchers may face when working with similar data.
Third, we derive practical guidance for ethically navigating such settings and, in doing so, distill four lessons for research on arXiv and similar public, live datasets: public availability should serve as the starting point of ethical reasoning; research should minimize the amplification it induces on exposed information; disclosure should be calibrated to the harm, confidence, and actionability of a finding; and effective immutability must be accounted for.
Together, these lessons contribute to community resources for ethically-sound cybersecurity experimentation.
\end{abstract}

%% file: paper_content/1_intro.tex
\section{Introduction} %
\label{sec:introduction}

Cybersecurity research often touches on issues that raise ethical questions~\cite{Kohnoetal2023Ethical}.
Beyond empirical studies, researching and disclosing vulnerabilities arguably also benefits attackers who can use the information to compromise vulnerable systems.
To that end, responsible disclosure practices have been established in the field~\cite{Householderetal2017The} and are widely used~\cite{Walsheetal2020An}.
In contrast, public datasets used in cybersecurity research can come with different trade-offs and warrant careful ethical considerations.
Passwords leaked by hackers after data breaches are used in research, \eg{}~\cite{Gollaetal2018On}, despite profiting from malicious actors~\cite{Kohnoetal2023Ethical}.
Similarly, data on Reddit~\cite{Proferesetal2021Studying} or GitHub~\cite{Melietal2019How}, for example, is widely used in research, even though individuals contributing to these live datasets may not be aware that their data is used for research~\cite{Shiltonetal2021Excavating,Shahriarietal2026A} and that their consent is rarely, if ever, obtained~\cite{Zimmer2018Addressing,Fiesler2019Ethical}.

The arXiv preprint repository~\cite{arXiv1991arXiv.org} represents one such public dataset and is an especially complex ethical case.
On arXiv, researchers publish their work under their real names, resulting in trivial author identifiability.
Prior work~\cite{Kurtzetal2005The,Davisetal2007Does,Moed2007The,Lariviereetal2014arXiv,Haqueetal2009Positional,Haqueetal2010Last,Okamura2022Scientometric} extensively sourced this resource to understand citation dynamics, for example.
Another peculiarity of arXiv is that authors must also upload their \LaTeX{} source files~\cite{arXiv2005Requiring} when posting preprints to foster research transparency.
Yet authors might not be aware that the paper sources they upload will be fully openly available, including all comments and other artifacts they might have forgotten to delete before uploading.
At the same time, arXiv intentionally does not permit the correction of information disclosed by accident in such source files~\cite{Apruzzeseetal2026X-raying,Pennekampetal2026Hidden,Dubniczkyetal2026You}.
Despite this situation, a growing body of literature has used these sources for empirical studies.

The immediate implications are twofold.
First, due to its increasing popularity as a dataset~\cite{Apruzzeseetal2026X-raying,Pennekampetal2026Hidden,Dubniczkyetal2026You}, associated ethical challenges are becoming more widespread. %
Second, given its popularity~\cite{arXiv2014Monthly-Downloads} and market share~\cite{Xieetal2021Is}, many researchers are (unknowingly) affected by these studies and their practices.
In turn, these observations motivate our assessment.

\textbf{Paper Story.}
Based on our experience of working with paper sources that arXiv serves publicly, an ethics-focused analysis of existing work~\cite{Apruzzeseetal2026X-raying,Pennekampetal2026Hidden,Dubniczkyetal2026You}, and discussions with affected authors, we map out the landscape of ethical issues and considerations surrounding the use of this data.
Specifically, in \cref{sec:background}, we describe the background and related work, and, in \cref{sec:problem}, we then outline the unique properties of the arXiv repository as a dataset.
Continuing with \cref{sec:decisions}, we detail how prior empirical studies approached the emerging ethical aspects.
Based on this assessment, in \cref{sec:dilemmas}, we subsequently derive ethical dilemmas that can arise in research using similar data.
Finally, we highlight important takeaways and remaining challenges in \cref{sec:discussion} before concluding this experience paper in \cref{sec:conclusion}.

\textbf{Contributions.}
In particular, we make the following primary contributions to ethical cybersecurity experimentation:
\begin{itemize}[leftmargin=*]
	\item \textbf{Mapping the ethical setting:}
		We describe the unique properties of arXiv and map the ethical issues that can arise in research that uses its data.
		This setting outlines how even data intentionally made public with real-name attribution can raise ethical concerns when used in research.
		In doing so, we contribute to a better understanding of how to work ethically with live resources.
	\item \textbf{Identifying dilemmas:}
		Rooted in our discussion of prior work and their ethical decisions, we abstract the ethical issues surrounding the use of arXiv data into four distinct dilemmas that researchers might face when using similar data sources, supporting future decision-making.
	\item \textbf{Guidance:}
		We debate the ethical responsibilities of researchers and provide practical guidance on navigating the issues that can arise.
		This way, we provide broader lessons for researchers working with arXiv and similar datasets.
\end{itemize}

%% file: paper_content/2_background.tex
\section{Background and Related Work} %
\label{sec:background}

In \cref{sec:background:arxiv}, we first introduce the arXiv repository and research that utilizes source files hosted there.
Subsequently, in \cref{sec:background:ethics}, we shift our focus to ethical guidance most relevant to analyzing arXiv-like knowledge infrastructures.

\subsection{arXiv as a Live Knowledge Infrastructure}
\label{sec:background:arxiv}

Since its inception in 1991, arXiv~\cite{arXiv1991arXiv.org} has become a dominant open-access preprint service, now holding over \qty{3}{\million} submissions and serving more than \qty{60}{\million} connections per day, with computer science accounting for the largest share of preprints added each month~\cite{Apruzzeseetal2026X-raying,Pennekampetal2026Hidden}.
Rooted in a principle of openness, arXiv also distributes the \LaTeX{} sources behind roughly \qty{90}{\percent} of submissions that take its recommended source-based route, originally to enable the reuse of notation, and today also framed as supporting transparency and long-term reproducibility.
It further offers this corpus for bulk retrieval via Amazon S3~\cite{Apruzzeseetal2026X-raying,Pennekampetal2026Hidden,Dubniczkyetal2026You,arXiv2010Full}.
Authors, in turn, grant a ``perpetual'' and ``irrevocable'' distribution license, while arXiv's privacy policy places the burden of removing anything not intended for publication on the submitter~\cite{Dubniczkyetal2026You}.

While preprint adoption is difficult to quantify, studies paint a similar picture:
Even back in 2017, specific subfields saw about \qty{23}{\percent} of papers on arXiv.
Another study reports that \qty{6}{\percent} of \emph{all} publications across disciplines were available on arXiv in 2021~\cite{Xieetal2021Is}, with physics, maths, and computer science above that average.
Specifically in computer science, \qty{77}{\percent} of posted preprints between 2008 and 2017 were eventually published~\cite{Linetal2020How}. %
To put the exposure to the scientific community in numerical terms, already by 2016, \qty{53}{\percent} of arXiv users were themselves authors~\cite{Riegeretal2016arXiv}.
Borgman~\cite{Borgman2017Big} characterizes such systems as \emph{knowledge infrastructures} that evolve over time and, in doing so, shape scientific practice.
Prior work mostly treats arXiv as a fixed snapshot (\cf{} \cref{sec:problem}), disregarding indirect consequences:
Findings on such knowledge infrastructures can directly shape and impact the science landscape and its actors.

Three recent studies analyze this corpus for security and privacy issues.
Apruzzese and Fass~\cite{Apruzzeseetal2026X-raying} report that across the source files of \qty{600}{\thousand} submissions, \qty{27}{\percent} of the uploaded data is unnecessary for LaTeX compilation, including offensive text and details of ongoing research.
Similarly, Pennekamp et al.~\cite{Pennekampetal2026Hidden} discovered credentials, private keys, and links to third-party resources after scanning the entire corpus, and propose a sanitization tool as a mitigation.
Lastly, Dubniczky et al.~\cite{Dubniczkyetal2026You} apply large language models to detect information leakage across \qty{100}{\thousand} submissions.
All three include ethics statements yet reach different conclusions on the same corpus, \eg{} on whether to release tooling, and on whether following embedded links is acceptable.
These different perspectives on an unambiguously public dataset motivate our work.

\subsection{Ethical Guidance for Research on Public Data}
\label{sec:background:ethics}

Given that the arXiv corpus spans multiple dimensions, covering both contained and referenced data, several ethical considerations are also important.
Next up, we will thus feature those directions that are generally of interest. %

\textbf{Ethics Frameworks.}
The Association of Internet Researchers (AoIR) guidelines are an established reference for Internet research ethics.
Version~1.0~\cite{Essetal2002Ethical} is deliberately pluralistic and case-based; Version~2.0~\cite{Markhametal2012Ethical} adds three commitments with implications to our setting:
\begin{enumerate*}[label=(\roman*)]
	\item greater vulnerability of the studied party implies a greater obligation of the researcher;
	\item ``human subject'' as a condition fits text-based research poorly and one should instead consider caused harm, vulnerability, and identifiability; and
	\item public/private boundaries are not binary, as people act in public venues while retaining expectations about how their data flows.
\end{enumerate*}
Version~3.0~\cite{Franzkeetal2020Internet} extends them to Big Data and machine learning, introduces a taxonomy spanning design through dissemination and project close, and poses dedicated questions for \emph{existing datasets}: whether a dataset is explicitly open for research use, whether any initial consent covers the intended reuse, and whether data subjects could conceivably object to it.
Its accompanying impact model separates the immediate treatment of people from side effects and from downstream use of data beyond the initial analysis.

Complementarily, Metcalf and Crawford~\cite{Metcalfetal2016Where} argue that data science is frequently research \emph{about} humans without qualifying as human-subjects research, so the regulatory exemption for publicly-available data is based on a superseded assumption.
Nissenbaum's contextual integrity~\cite{Nissenbaum2009Privacy} supplies an alternative, framing privacy as appropriate information depending on context.
Zimmer~\cite{Zimmer2018Addressing} specifies a framework for Big Data that addresses concerns around consent, anonymity, and harm that a binary public/private test leaves unaddressed.

Data statements~\cite{Benderetal2018Data}, datasheets~\cite{Gebruetal2021Datasheets}, and accountability practices adapted from software engineering~\cite{Hutchinsonetal2021Towards} have also been proposed.
All assume a curated artifact with an identifiable creator who can state provenance, composition, and intended use, which, however, is not the case for arXiv.

\textbf{Empirical Practice.}
Surveying \num{263} online-data researchers, Vitak et al.~\cite{Vitaketal2016Beyond} find that disagreements about ethical practices are uncorrelated with discipline, method, or workplace.
Their follow-up with staff at \num{59} US institutional review boards (IRBs)~\cite{Vitaketal2017Ethics} finds no consensus on which social computing protocols require review, alongside a self-reported lack of technical expertise.
Bruckman~\cite{Bruckman2002Studying} models Internet users as amateur artists producing ``semi-published'' work and concludes that the resulting material has to be treated carefully.
Regarding ethicality, they argue that being too lax risks violating individual subjects’ rights and disrupting the community, while strictness might push researchers to avoid IRB oversight in the first place.

Data subjects, when asked, seem to not generally endorse a public-data exemption:
\qty{61.2}{\percent} of surveyed Twitter users did not know that researchers use tweets at all, and \qty{64.9}{\percent} noted that researchers should not do so without permission, with comfort dropping sharply if entire histories instead of isolated posts are studied~\cite{Fiesleretal2018Participant}, contrasting IRB-centered studies that report IRB respondents feel that researchers do not need informed consent for public data~\cite{Vitaketal2017Ethics,Fiesler2019Ethical}.
Comparable expectations hold in online fandom~\cite{Dymetal2020Ethical}.
Similarly, users accept population-level depression monitoring far more readily than individual-level assessment~\cite{Mikaletal2016Ethical}, suggesting that aggregation may be a potential mitigation.
Reported practice, nonetheless, is scarce:
Across \num{727} Reddit studies, only \qty{13.9}{\percent} mention any ethics review, mostly to record an exemption.
\qty{6.1}{\percent} mention consent, \qty{9.4}{\percent} name identifiable usernames, and \qty{28.5}{\percent} quote users verbatim~\cite{Proferesetal2021Studying}.

\textbf{Internet Measurement Ethics.}
Network measurement has developed its own norms, centered on the Menlo Report~\cite{Dittrichetal2012The} and on scanning practices such as rate limiting, attributable probes, and opt-out~\cite{Durumericetal2013ZMap:}.
Yet Fiebig~\cite{Fiebig2023Crisis} argues that this practice systematically undermines both the ethics and the reliability, and Amin et al.~\cite{Aminetal2025Navigating}, based on \num{16} interviews, report that researchers treat ethics as craft knowledge passed on by mentors and peers, because review boards lack the technical grounding to assess such work.
Studies of exposed secrets on the web~\cite{Dahlmannsetal2023Secrets,Demiretal2026Keys} come closest to analyzing arXiv submissions in subject, but their data subjects are usually service operators and only rarely named individuals.

Closest to our work, Kohno et al.~\cite{Kohnoetal2023Ethical} structure security-specific dilemmas as trolley problems to make them explicit. %
Their scenarios, like most work on dataset ethics, concern---by design---data deliberately released by a third party with foreseeable negative implications, or data obtained illegitimately (\eg{} stolen or leaked).
However, arXiv source files are neither:
They were deliberately released by the affected individuals themselves under a norm of openness and without any expectation that residual data would be read.

%% file: paper_content/3_problem.tex
\section{Problem Statement: arXiv as a Live Dataset} %
\label{sec:problem}

From an abstract perspective, the arXiv ``dataset'' is comparable to other live resources that are available on the Internet, such as social media content or snapshots from Internet scans (\eg{} Censys~\cite{Censys2017Censys} or Shodan~\cite{Shodan2013Shodan}).
Because these resources continue to evolve, they can in principle support longitudinal monitoring of individuals, including observations of newly contributed data as it becomes publicly available.
They all have in common that they are not released, curated datasets (no official identifier/reference), but living publicly-available datasets.
Further, they neither disclose to affected data subjects that they are part of the dataset nor do they offer effective means to opt out of studies, since the compiled data is publicly available.
In health research involving identifiable patient data, for example, access and reuse are commonly governed by formal consent, authorization, or waiver mechanisms~\cite{Ballantyneetal2018Consent,Department-of-Health-and-Human-Services2018U.S.}.
However, as we discuss in \cref{sec:problem:arxiv}, arXiv as a dataset comes with unique aspects; hence, established ethical guidelines are not comprehensive enough.
Our assessment of related work in \cref{sec:problem:ethics} further underlines that lacking or superficial ethical statements are the norm rather than an exception, calling for a behavior change.

\subsection{arXiv's Uniqueness and Dataset Dimensions}
\label{sec:problem:arxiv}

Different from the other live resources we have just discussed, the arXiv dataset comes with three unique aspects:

\emph{$\blacktriangleright$ Intentional Behavior.}
Unlike in many measurement settings, where individuals become part of a dataset incidentally or without taking deliberate action, arXiv authors intentionally submit their papers and associated source files for dissemination.
However, intentionally submitting a paper for dissemination does not necessarily imply awareness of, or intent regarding, every piece of information they contain, \eg{} when considering its associated source files.
Authors may intend to make their paper publicly available while remaining unaware that residual comments, obsolete text, or other source artifacts are also publicly retrievable, as suggested by prior work~\cite{Apruzzeseetal2026X-raying,Pennekampetal2026Hidden}.

\emph{$\blacktriangleright$ Strong Identifiability.}
All arXiv submissions and their source files are, by design, directly associated with a fixed set of named authors.
This differs from many other datasets in which individuals are anonymous, pseudonymous, or identifiable only through additional inference.
Consequently, findings derived from a submission can readily be associated with identifiable researchers or research teams, even when the authorship of a particular source-file artifact cannot be established.
Such identifiability may therefore increase the reputational or professional implications of sensitive findings.

\emph{$\blacktriangleright$ Effective Immutability.}
While arXiv, as a service and dataset, is not per se immutable, its operation effectively results in established immutability.
Authors have little means to remove or sanitize sensitive information from the dataset, as arXiv opts for amendments and updates rather than replacing data, thereby circumventing permanent deletion.

Moreover, affected individuals are not merely data subjects: they may simultaneously be authors of the work, contributors to the submitted source files, submitters responsible for the arXiv record, and private individuals.
As a result, ethical assessments may differ across the roles they affect.

\textbf{``Similar'' Datasets.}
Researchers frequently use GitHub~\cite{GitHub-Inc.2008GitHub}, the Wayback Machine~\cite{Internet-Archive2001Wayback}, and Zenodo~\cite{European-Organization-For-Nuclear-Researchetal2013Zenodo} as datasets for studies~\cite{Aroraetal2016Using,Cosentinoetal2016Findings,Melietal2019How,Petersetal2017Zenodo,Siciliaetal2017Community,Sandtetal2019Practice,Gargetal2025A,Schreiberetal2025Security,Zarrilloetal2025Web}.
All of them match a subset of arXiv's unique aspects:
GitHub typically serves author-provided data that includes a history and provides strong identifiability, often including embedded author email addresses.
In contrast to arXiv, pseudonyms are officially supported, and hosted repositories are not immutable.
Besides, different visibility settings enable access control.
On a different note, the Wayback Machine also provides an immutable dataset with version history.
However, unlike arXiv, the recorded data does not necessarily provide strong identifiability, and information is only available in the exceptions that the authors themselves provide.
Lastly, Zenodo most closely matches arXiv's aspects; however, authors can also be organizations, slightly relaxing the strong identifiability.

\textbf{Dataset Dimensions.}
Source files served by arXiv---\emph{the dataset}---specifically touch upon three distinct dimensions.

\emph{$\blacktriangleright$ \internal{}.} %
This dimension covers information confined to the dataset itself, primarily conversational \LaTeX{} comments (\eg{} author conversations), as well as todo lists and obsolete words, sentences, and paragraphs.
Other examples include details on submission histories or (prior) template use, reviews and revisions letters, or use of profanity.

\emph{$\blacktriangleright$ \mixed{}.} %
Data belonging to the second dimension links information from the dataset with third-party services, \ie{} only in combination is the data of value.
The most prominent example covers access tokens or passwords that can be used to authenticate users.
Other, less straightforward cases are identifiers (\eg{} email or cryptocurrency addresses and paper submission IDs), credit card details, or GPS coordinates.

\emph{$\blacktriangleright$ \external{}.} %
Finally, the third dimension captures data that solely points to external services.
Common examples include links to cloud services (including Google Docs or Drive), FTP servers, and IP addresses.
This information is not actionable within the dataset itself; it is independent and largely exclusive to the referenced service.

We do not find a strict mapping between these dimensions and the different types of ``hidden'' information.
That is, each dimension touches on the file, metadata, and content/text levels~\cite{Pennekampetal2026Hidden}; however, the \emph{\mixed{}} dimension is particularly prominent as part of dangling files~\cite{Pennekampetal2026Hidden}.
Lastly, these dimensions (\cref{tab:dimensions}) mandate individual ethical considerations to account for the diversity in sensitivity, potential harm, and affected stakeholders.

\subsection{Ethics Statements in Related Work Using arXiv}
\label{sec:problem:ethics}

In the context of using arXiv as a dataset, the research landscape reveals that most studies do not discuss ethical implications at all.
Overall, we come across various research directions, working with \LaTeX{}~\cite{Cormodeetal2012Scienceography:,Cormodeetal2013Studying}, quantifying the impact of preprints~\cite{Kurtzetal2005The,Davisetal2007Does,Moed2007The,Haqueetal2009Positional,Haqueetal2010Last,Lariviereetal2014arXiv}, sourcing content from papers hosted at arXiv~\cite{Daweietal2020A,Ginevetal2020Scientific,Lietal2020DocBank:,Scharpfetal2020Classification,Meuschkeetal2023A}, interacting with served papers~\cite{Sorokinaetal2006Plagiarism,Citronetal2015Patterns,Safnuketal2018Reconstructing,Akram2024Quantitative}, and understanding authorship and publication forms~\cite{Clementetal2019On,Saieretal2020unarXive:,Bianchi2025Text,Hepler2025Modular,Peietal2025Hidden,Yangetal2025Writing}.
This situation is particularly surprising since some works~\cite{Sorokinaetal2006Plagiarism,Citronetal2015Patterns,Akram2024Quantitative,Peietal2025Hidden,Yangetal2025Writing} study arguably sensitive matters like plagiarism detection and scientific misconduct, or publish arXiv-extracted datasets themselves~\cite{Saieretal2020unarXive:}.
Interestingly, all papers investigating arXiv source files for secrets~\cite{Apruzzeseetal2026X-raying,Pennekampetal2026Hidden,Dubniczkyetal2026You} feature ethics statements of varying length (\num{900}, \num{1098}, and \num{420} words, respectively), distinguishing the decisions to conduct the research and to publish the findings.
To the best of our knowledge, current assessments are limited to selected papers~\cite{Raoetal2024WithdrarXiv:,Apruzzeseetal2026X-raying,Pennekampetal2026Hidden,Dubniczkyetal2026You} that indeed discuss privacy implications.

We further looked at \num{35} papers that utilize the \emph{ogb(n)-arxiv}~\cite{Open-Graph-Benchmark2020Dataset,Huetal2020Open} dataset, which contains the citation network of computer science research, originally extracted from papers published at arXiv.
Most of them omit an ethics section entirely; three papers disregard negative consequences, albeit two without discussing privacy. %
The remaining paper argues that no new data collection took place (\cf{} \cref{sec:background:ethics}). %

\textbf{Research Gap.}
In our view, the nuances of the arXiv ecosystem and their intersection with ethics are not yet sufficiently covered by prior work, warranting a closer look.
In fact, research areas differ in their stance toward ethics:
While studies working with social media content largely argue that the data is public information and thus their work is exempt from data subject consent and IRB approval~\cite{Proferesetal2021Studying}, Internet measurements usually discuss their ethical implications (\cf{} \cref{sec:background:ethics}) but do not take specific action when solely relying on readily-available datasets from by third-party services, such as Censys or Shodan, \ie{} they also conclude that they are not directly interacting with the data subject.

%% file: paper_content/4_assessing_ethics.tex
\section{Assessing Ethics Decisions in Prior Work} %
\label{sec:decisions}

The ethics statements of prior work~\cite{Apruzzeseetal2026X-raying,Pennekampetal2026Hidden,Dubniczkyetal2026You} are the focus of our assessment.
In line with this paper's scope (\cf{} \cref{sec:introduction}), we limit ourselves to studies that work with paper sources, \ie{} that empirically analyze the arXiv dataset.
In particular, we highlight the reported ethical considerations and decision-making, and further discuss the corresponding outcomes to understand the implications of these decisions.

To this end, in \cref{sec:decisions:infrastructure}, we first briefly examine how the studies affect arXiv's operation before individually covering the three dataset dimensions in \cref{sec:decisions:dimensions}.
Finally, we conclude our assessment in \cref{sec:decisions:immutability} by focusing on how the authors account for arXiv's effective immutability.

\subsection{Interacting with a Live Service}
\label{sec:decisions:infrastructure}

Given that arXiv is a service used by millions of users daily~\cite{arXiv2014Monthly-Downloads}, ethical research must ensure it does not disrupt the service.
Compliant with arXiv's recommendation for bulk access~\cite{arXiv2010Full}, prior work almost exclusively relied on the official Amazon S3 endpoint, rather than directly interacting with the production system's infrastructure.
Our interpretation is that researchers likely avoided straining the service's resources, which could otherwise be an ethical concern.
Only Pennekamp et al.~\cite{Pennekampetal2026Hidden} accessed the official arXiv API to complete their dataset.
Nevertheless, they reported that they remained within the documented, appropriate usage rates~\cite{arXiv2019Terms}.
Collectively, the decisions made are in line with other analyses, \eg{} of citation graphs (\cf{} \cref{sec:problem:ethics}).
Thus, the research-induced impact on arXiv and its operation appears to have been marginal in all prior studies, not raising deeper-rooted ethical issues.
In the following, we shift our attention to more challenging matters.

\subsection{arXiv's Dataset Dimensions and Implications}
\label{sec:decisions:dimensions}

\begin{table}
	\centering
	\caption{%
		Examples of the dataset dimensions introduced in \cref{sec:problem:arxiv}, along with their retrospective remediability and potential sensitivity.
	}
	\label{tab:dimensions}
	\vspace{-0.5em}

	\begin{scriptsize}
	\renewcommand{\arraystretch}{1.12}
	\setlength{\tabcolsep}{2.25pt}

	\begin{tabularx}{\columnwidth}{
		>{\raggedright\arraybackslash}p{1.8cm}
		>{\centering\arraybackslash}p{1.25cm}
		>{\centering\arraybackslash}p{1.35cm}
		>{\raggedright\arraybackslash}X
	}
	\toprule
	\textbf{Dimension} &
	\textbf{Remediability} &
	\textbf{Sensitivity} &
	\textbf{Representative Examples} \\
	\midrule

	\internal{} &
	Low / \xmark{} &
	\impL{}--\impM{} &
	Author conversations, profanity, in-file todo lists, obsolete text and markup, LLM response patterns \\

	\addlinespace[2pt]

	\mixed{} &
	Partial / \smark{}&
	\impM{}--\impH{} &
	Access tokens, credentials, private keys, (user) identifiers, credit card details, GPS coordinates \\

	\addlinespace[2pt]

	\external{} &
	High / \cmark{} &
	\impL{}--\impH{} &
	Links to cloud services, websites, file servers, IP addresses \\
	
	\bottomrule
	\end{tabularx}

	\vspace{2pt}
	Remediability describes how readily an issue can be addressed retrospectively, given arXiv's effective immutability (\cf{} \cref{sec:problem:arxiv}).
	Sensitivity can only be classified subjectively; in line with prior work~\cite{Leonetal2013What,Pennekampetal2026Hidden}, we assign high~(\impH{}), medium~(\impM{}), or low~(\impL{}) according to potential impact following public exposure of information.
	\end{scriptsize}
	\vspace{-1.0em}
\end{table}

The identified dataset dimensions (\cref{sec:problem:arxiv}) also come with unique implications.
\cref{tab:dimensions} provides an overview of the different dimensions and whether discovered issues are fixable, given arXiv's effective immutability.

As prior work~\cite{Pennekampetal2026Hidden}, we \emph{coarsely} label findings as high (\impH{}), medium (\impM{}), or low (\impL{}) to indicate their \emph{potential} severity.
Although such a labeling is highly subjective~\cite{Pennekampetal2026Hidden}, a coarse categorization remains useful for distinguishing the relative severity of different findings and structuring the subsequent ethical assessment.
The examples detail which findings \textit{may} trigger negative implications.
While the exposure of access tokens or links to publicly accessible cloud services (without enforced access control) is simple to grasp, other aspects are less trivial.
Aspects such as using profanity, bashing competitors, or ``exposing'' scientific misconduct can be career-altering.
Similarly, GPS coordinates, address details, or the use of GeoIP databases in combination with IP addresses can have real-world implications, potentially leading to in-person stalking.
In particular, since authors cannot address all issues retrospectively, researchers must carefully consider the (real-world) implications when conducting their analyses.

Given their differences, we now assess the three dimensions commonly featured in paper sources on arXiv, step by step, and outline the corresponding ethical considerations.

\textbf{\internal{}.}
The combination of the scale (millions of unique comments) and context-dependent interpretation (obsolete or ``only'' uncommented remarks) generally impairs accurate identification of the respective issues with certainty.
Notwithstanding, author conversations, obsolete text fragments, or outdated comments have mostly local implications.
Likely, given this context dependence and the limited ability to remediate identified instances retrospectively, prior work has reported on the general prevalence rather than highlighting individual cases.
As a safety measure, prior work collectively refrained from pointing to individual instances (and their identifiers), retaining the needle in the haystack.
Due to these factors, we perceive \internal{} to be low sensitivity (\impL{}).
Still, scientific misconduct could even be career-altering, contributing to higher sensitivity (\impM{}; possibly even \impH{} in extreme cases).

\textbf{\mixed{}.}
Here, all studies decided against ``testing'' any authentication tokens or credentials, which matches with prior work that discovered ``hidden'' secrets~\cite{Melietal2019How}.
This approach appears to be in line with the content the Menlo report~\cite{Dittrichetal2012The} expresses~\cite{Dubniczkyetal2026You}.
The same holds for (user) identifiers, credit card details, or information that potentially reveals the author(s)' whereabouts.
Interestingly, unlike the studies we cover, recent work~\cite{El-Yadmanietal2025The,Demiretal2026Keys} instead decided to verify their findings by ``unintrusively'' connecting to affected services.
Still, we posit that connecting to third-party services greatly increases the risk of harm and may also unintentionally trigger safety guards (which is positive as a responsible disclosure mechanism) or notifications (which may be negative, as they can stress affected/notified users).

\textbf{\external{}.}
This externally-focused dimension is very closely related to Internet measurements and their guidelines (\cf{} \cref{sec:background:ethics}).
Given the diversity of findings and referenced services, servers, and documents, processing these pointers touches upon a broad(er) set of external endpoints.
Unlike the credentials and tokens reported previously, two studies~\cite{Apruzzeseetal2026X-raying,Pennekampetal2026Hidden} accessed Google Docs links to determine whether authors had enabled access control and, if not, possibly to understand what kind of data is exposed in this way.
Their decisions to restrict themselves to a subset minimize harm and avoid the use of exposed credentials. %

Users accessing the source files usually have no way to determine whether accessing a referenced resource is ``unauthorized'' or not.
While ``hidden'' references are not part of the compiled PDF, it is fair to assume that authors are unlikely to annotate the sensitivity of these references either.
The remaining risks are two-fold.
First, researchers may access sensitive information that was not intended for public access.
Second, researchers may unwillingly be exposed to illegal or disturbing content.
Still, the benefits of learning what kind of information may be (unwillingly) exposed in a small sample outweigh the arguably marginal harm to affected authors, particularly because malicious actors---with no intent to inform affected authors---could easily do so as well.
The discoveries vary greatly in sensitivity from \impL{} to \impH{}, which we attribute to the diversity of referenced content.
However, this decision of sensitivity is disputed among prior work~\cite{Apruzzeseetal2026X-raying,Pennekampetal2026Hidden,Dubniczkyetal2026You} and should be reevaluated frequently.

Better understanding which other resources %
and with which permissions are exposed could further characterize the scope and nature of these external references.
However, we expect that such an extended validation would reveal similar findings at greatly amplified risks, given potentially less predictable service behavior.
Most likely, such attempts would even again reveal transitively-exposed leaks~\cite{Pennekampetal2026Hidden}.

\subsection{Accounting for arXiv's Effective Immutability}
\label{sec:decisions:immutability}

As summarized in Table~\ref{tab:dimensions}, the three dataset dimensions differ not only in their potential sensitivity but also in their remediability.
These differences are closely tied to arXiv's \emph{effective immutability}, a distinguishing property of the dataset.
Next, we examine how prior work navigates this constraint.

\textbf{Immutability Challenge.}
Given that arXiv is effectively immutable and arXiv argues that authors are responsible~\cite{Pennekampetal2026Hidden}, mitigation strategies to combat ``hidden'' information are quite limited.
Certainly, for \emph{\external{}}, authors can likely (re-)configure access control and authorization directly at the exposed third-party service, leading to comparably good remediability (\cmark{}).
Similarly, in terms of \emph{\mixed{}}, certain tokens and credentials can be disabled or revoked, though they are not fully removed from the arXiv dataset.
Unfortunately, for other aspects, \eg{} included (user) identifiers, or location data, no such ``convenient'' option is available, resulting in only \smark{} remediability.
Lastly, any data of \emph{\internal{}} is not fixable at all (\xmark{}), simply because it is confined to the ``immutable'' data at arXiv.

While an (undocumented) removal of sources is possible in extreme cases, arXiv is not interested in broader action~\cite{Pennekampetal2026Hidden}.
As of 10/2026, we are in talks with arXiv to have them address a few severe issues.
Regardless, (un)authorized copies of the dataset result in data still floating, %
forcing affected authors to live with the fact that sensitive data is publicly exposed---in line with the mantra \emph{``The Internet never forgets''}.
Fortunately, actions like access-control adjustments or the reissuing of credentials leave some room for containment.

\textbf{Mitigation Support.}
Despite this somewhat grim outlook, two of the three papers~\cite{Apruzzeseetal2026X-raying,Pennekampetal2026Hidden} initiated a responsible disclosure %
to inform %
affected authors.
Here, a key question is whether this outreach is ethically warranted
\begin{enumerate*}[label=(\alph*)]
	\item given that many issues are not fixable and
	\item sensitivity assessments of issues are highly subjective,
\end{enumerate*}
possibly resulting in stressful and undesired situations.
Given that the dataset remains publicly accessible, we conclude that the early warning paired with the option to apply some fixes (\cmark{} and \smark{}) is still beneficial and probably in the broader interest of affected authors.

%% file: paper_content/5_touched_ethical_dilemmas.tex
\section{Touched Ethical Dilemmas}
\label{sec:dilemmas}

After detailing the ethical considerations that affected the aforementioned arXiv studies, we now take a step back and abstract the reported situation into higher-level dilemmas.
That is, %
we derive dilemmas that may also apply to other research that involves live datasets.
Specifically, we want to raise awareness by highlighting four dilemmas that warrant careful evaluation in future studies.
While the first two dilemmas touch upon the accuracy of the research, namely quantifying and validating findings, the third dilemma concerns the scale of empirical studies.
Lastly, we turn to a fourth dilemma that concerns the responsible disclosure of research findings and the associated benefits, or lack thereof.

\begin{dilemmablock}{Sensitivity vs.\ Need for Quantifying}{dilemma:sensitivity} %
\end{dilemmablock}

The \emph{ACM Code of Ethics}~\cite{ACM2018ACM} calls for respecting privacy, yet quantifying how much sensitive information a dataset exposes, and how severe it is, requires classifying that information, which in turn requires reading it.
This dilemma thus concerns the tension between accessing potentially sensitive private information and the need to quantify its prevalence and sensitivity.
The dilemma itself persists across studies, whereas a legitimate research interest is a case-specific factor that may resolve it:
For example, establishing that exposed links indeed lead to unprotected documents requires opening a sample of them, while counting such links avoids storing private information, but the resulting estimates may significantly over- or underreport the actual exposure.
In the studies we assessed, all three~\cite{Apruzzeseetal2026X-raying,Pennekampetal2026Hidden,Dubniczkyetal2026You} processed \emph{\internal{}} at scale to report prevalence but published aggregates only; for \emph{\external{}}, two~\cite{Apruzzeseetal2026X-raying,Pennekampetal2026Hidden} opened a small sample of referenced documents to assess sensitivity (\cf{} \cref{sec:decisions:dimensions}), while the third~\cite{Dubniczkyetal2026You} did not.

\begin{dilemmablock}{Validation Attempts vs.\ Overreporting}{dilemma:validation} %
\end{dilemmablock}

The second dilemma is closely related and boils down to intentionally ``working'' with credentials or other data to validate their authenticity, \eg{} for authorization purposes or to delimit access boundaries.
Different views have emerged (\cf{} \cref{sec:decisions:dimensions}), with some~\cite{El-Yadmanietal2025The,Demiretal2026Keys} arguing that careful validation attempts are key to avoiding excessive reporting of false positives.
We certainly acknowledge the risk of overreporting findings but generally believe that accessing services and resources using data meant to remain private without consent usually contradicts desirable norms~\cite{ACM2018ACM,Dittrichetal2012The}.
Across the three studies we investigated, none attempted to validate user credentials, access tokens, or credit card information, to our understanding.
Instead, Pennekamp et al.~\cite{Pennekampetal2026Hidden} confined their validation to non-invasive approaches, such as checking for valid semantic patterns or recalculating embedded checksums, to assess how authentic they appear.
The other two studies~\cite{Apruzzeseetal2026X-raying,Dubniczkyetal2026You}, in contrast, did not report corresponding validation attempts in their papers.

\begin{dilemmablock}{Problem Scale vs.\ Available Resources}{dilemma:scale} %
\end{dilemmablock}

Particularly when dealing with massive datasets, research is challenged by the scale of the problem and may sacrifice the usual precision and extensive coverage for two reasons.
First, if (fully) automated processing is not possible, for example, because human reasoning or contextualization is required, the analysis can quickly exceed the available ``processing'' resources.
Likewise, the human attention span is quite limited~\cite{Langneretal2013Sustaining}, which may lead to unintentional bias or mental overload for the conducting researchers.
Second, from an ethical perspective, sustainability considerations may also play a key role, \eg{} the desire to avoid excessively ``burning'' computing power~\cite{Lannelongueetal2021Ten}---an aspect that is increasingly relevant in the age of large language models (LLMs).
While this dilemma is certainly not confined to the arXiv, prior work reveals significant differences in how it is addressed.
On the one hand, the considered dataset sizes vary significantly, as we introduced in \cref{sec:background:arxiv}.
On the other hand, resource utilization also differs, ranging from no reported use of LLMs~\cite{Apruzzeseetal2026X-raying}, through selected use~\cite{Pennekampetal2026Hidden}, to work that places LLMs at the center of a research question~\cite{Dubniczkyetal2026You}.
While a direct comparison of the environmental footprint is not possible, this overview highlights that researchers come to different conclusions concerning what is sustainable.

\begin{dilemmablock}{Awareness Benefit vs.\ Attention}{dilemma:awareness} %
\end{dilemmablock}

Moving to a human-centered dilemma, responsible disclosure intends to raise awareness among affected individuals, but the benefits may not always outweigh the harms it can cause.
Li et al.~\cite{Lietal2016Youve} argue that when individuals are notified, the potential good outweighs the potential harm.
Still, learning that one's data was part of a data breach can evoke a range of negative emotions, from concern~\cite{Mayeretal2021Now,Mayeretal2023Awareness} to severe emotional distress~\cite{Soloveetal2017Risk}.
Thus, when informing affected stakeholders, researchers should also consider how beneficial the resulting awareness is, particularly when weighed against the stress and time expenditure that a notification may cause.
Past research~\cite{University2021Princeton-Radboud} shows that even good intentions can have negative consequences. %
The wording of the notification can also be essential: legal framings of notification messages can be effective in eliciting a response~\cite{Maassetal2021Effective}, but may also cause severe reactions~\cite{Hennigetal2025I}.
Compounding this issue, poorly crafted notifications may be misclassified as spam by recipients~\cite{Hennigetal2025I}.

However, the benefits of raising awareness cannot be considered in isolation, as the same disclosure or publication activities may also direct additional attention toward the exposed information itself.
Research outputs, datasets, or analysis tools may make information that was previously difficult to discover substantially easier for third parties to locate, attribute, or act upon.
Pennekamp et al.~\cite{Pennekampetal2026Hidden}, for example, deliberately refrained from releasing their analysis code to prevent potential misuse and avoid enabling third parties to easily identify submissions containing sensitive information.
This decision illustrates how efforts to raise awareness of a problem can simultaneously increase the practical visibility of the underlying exposure.
Dubniczky et al.~\cite{Dubniczkyetal2026You}, in contrast, did not conduct a responsible disclosure.
This choice highlights the other side of the dilemma: foregoing notification avoids imposing additional burden on affected individuals and may limit further attention to the exposure, but it also forgoes any remediation benefit that awareness could provide.
However, the reasons for their decision are not reported, and we therefore cannot determine whether such considerations influenced it.

Consequently, while notifying affected individuals about identified threats or issues should generally be considered, the decision to do so must be a conscious one that weighs both the potential benefit and the potential harm.
For instance, in the case of arXiv, authors are not always able to remediate discovered issues while leftover comments might impact careers (\eg{} leftover harsh criticism of related work). 
This situation might lead to feelings of helplessness, resignation, or a sense of inevitability that cyber incidents will always happen and that one is at their mercy~\cite{Mayeretal2023Awareness}, which, in turn, may unnecessarily expose individuals to stress or lead them to spend time attempting fixes that are no longer possible.
At the same time, researchers should consider whether the manner in which they communicate or publish their findings increases the practical visibility of exposed information beyond what is necessary to achieve the intended awareness benefit.

%% file: paper_content/6_discussion.tex
\section{Discussion, Challenges, and Takeaways} %
\label{sec:discussion}

The previous sections illustrate that the ethical challenges surrounding arXiv do not arise simply because researchers encounter sensitive information.
Rather, they emerge from the interaction between public availability, strong identifiability, effective immutability and the possibility that research itself changes how exposed information can be discovered or acted upon.
In this section, we synthesize these observations into broader lessons for researchers working with arXiv and related ``immutable'' live datasets.
We deliberately address researchers rather than repository operators, as repository-side mitigations (\eg{} no longer serving source files publicly) have already been discussed in prior work~\cite{Apruzzeseetal2026X-raying,Pennekampetal2026Hidden,Dubniczkyetal2026You}.

\subsection{Public Availability and Ethical Responsibility}
\label{sec:discussion:responsibility}

Public availability should be the starting point of ethical reasoning rather than its conclusion.
Particularly in light of the increasing prevalence of open-source intelligence~\cite{Pastor-Galindoetal2020The}, information leaks pose increasing risks~\cite{Dubniczkyetal2026You}.
Prior work~\cite{Mikaletal2016Ethical,Fiesleretal2018Participant,Proferesetal2021Studying,Dymetal2020Ethical} on Twitter, Reddit, and other online platforms shows that technical accessibility does not necessarily imply that individuals expect their information to be collected or analyzed for research.
arXiv presents a nuanced case because authors intentionally submit their papers for public dissemination (\cf{} \cref{sec:problem}), but this intentional action does not necessarily imply equivalent awareness or expectations regarding how broadly the submission will be inspected.
For example, authors may post a preprint while it is under double-blind review without anticipating substantial readership at that stage~\cite{Pranavetal2026The}.
Moreover, authors may understand that the paper itself is public while remaining unaware that the accompanying source files are likewise publicly retrievable, as evidenced by survey respondents in Pennekamp et al.'s study~\cite{Pennekampetal2026Hidden}.
Residual artifacts within those sources, such as comments, TODOs, obsolete text, or internal discussion, may therefore not have been intended for broad inspection.
The resulting tension is therefore between open science and reproducibility, on the one hand, and privacy and data minimization, on the other.

Informed consent~\cite{Nijhawanetal2013Informed} provides limited practical guidance at this scale.
Researchers could theoretically request consent from individuals, %
but doing so across hundreds of thousands of submissions is unlikely to be feasible~\cite{Golderetal2017Attitudes} and would likely result in substantial non-response.
This situation also raises the familiar tension between opt-in and opt-out models for public-data research~\cite{Zimmer2018Addressing,Fiesler2019Ethical}.
Similarly, arXiv's scale and heterogeneity challenge identifying vulnerable populations in advance~\cite{Markhametal2012Ethical}.
Researchers should therefore remain attentive to whether particular findings, contexts, or groups introduce vulnerability as analyses progress.

These observations return us to the sensitivity-versus-quantification dilemma (\dilemmaref{dilemma:sensitivity}, \cref{sec:dilemmas}).
Rather than asking only whether information is publicly accessible, researchers should consider whether the proposed use is consistent with the context in which it was shared and whether inspecting more sensitive information is necessary to answer the research question.
Researchers should thus justify why the anticipated research benefit warrants the chosen level of inspection.
In particular, the potential harm to individuals, including feelings of helplessness or distress in the face of ``unfixable'' public disclosure, should be considered with care.

\subsection{Minimize Research-Induced Amplification}
\label{sec:discussion:amplification}

Researchers may not be responsible for creating the original exposure, but their work can amplify it.
Information hidden among millions of files differs from information that research makes easier to discover, validate, attribute, or redistribute.
The ethical question is therefore not only whether researchers may access information but also how their actions change its practical accessibility, particularly for actors with malicious intent, and impact.
This concern directly reflects the attention aspect of \dilemmaref{dilemma:awareness}: efforts to raise awareness of an exposure can simultaneously draw additional attention to the underlying information and make it easier for others to discover or act on.

Amplification can occur throughout the research process.
Automated discovery can transform ``needle in the haystack'' information into searchable findings.
Validation, such as following external references or interacting with credentials, can shift from passive observation to active intervention (\cf{} \dilemmaref{dilemma:validation}).
Publication can further amplify exposure when examples, datasets, or artifacts make sensitive findings easier to locate. %
At arXiv scale, automated processing is likely unavoidable given limited human resources (\cf{} \dilemmaref{dilemma:scale}), but false positives and missing context should constrain how confidently findings are interpreted and attributed.
In turn, research might even trigger real-world actions.

Particular caution is warranted for reputationally-sensitive findings.
A source-file comment stating that ``we fudged certain numbers'' may appear to indicate misconduct, yet researchers may not know who wrote it, whether it was serious or obsolete, or whether other authors were aware of it.
Since arXiv's strong identifiability links every source file to named authors (\cf{} \cref{sec:problem}), turning ambiguous evidence into an attributable claim risks creating a digital pillory.
Researchers should therefore favor aggregate reporting, avoid unnecessary identifiers, limit redistribution of sensitive examples, and constrain validation to what is scientifically necessary.
The greater the potential harm of a false positive, the stronger the evidence should be before attribution.

\subsection{Disclosure: Harm, Confidence, Actionability}
\label{sec:discussion:disclosure}

Responsible disclosure should depend on the nature of the finding:
Actionable security issues, such as exposed credentials or private keys, may warrant private notification to enable revocation, access-control changes, or other containment measures.
Conversely, ambiguous reputational findings, such as misconduct or sensitive author conversations, may cause substantial harm if misinterpreted or incorrectly attributed.

Disclosure activities can range from private notification or escalation to an appropriate intermediary or institution, to limited public attribution or non-disclosure.
The recipient of a notification is likewise not always evident, as affected individuals act simultaneously as authors, submitters, and private individuals (\cf{} \cref{sec:problem}), and the submitter of a paper need not be the person who wrote a sensitive comment.
Researchers must further consider that notification itself may also create stress when affected parties have little ability to remediate the underlying exposure (\cf{} \dilemmaref{dilemma:awareness}).
Decisions should therefore consider severity, confidence, who may be affected, false-positive risk, and whether recipients can meaningfully respond.
Thus, disclosure should be treated as a context-dependent harm-reduction decision.

\subsection{Immutability Influences the Ethical Calculus}
\label{sec:discussion:immutability}

Effective immutability, as prevalent for some live datasets, matters throughout the research lifecycle because findings differ in remediability: credentials or permissions may be changed.
Then again, comments or obsolete text may remain permanently accessible.
Research should therefore account for fixability when collecting, validating, and disclosing findings, favoring minimized collection, limited validation, and aggregate reporting when remediation is difficult.

Repositories may also consider narrowly-scoped remediation or redaction mechanisms for sensitive historical material, balancing privacy and harm reduction against provenance and reproducibility.
Similar concerns may arise for GitHub~\cite{Goldetal2022Ethics}, the Wayback Machine~\cite{Hodges2022How}, or Zenodo, where contributors may not expect technically public material to be systematically examined.
However, differences in platform purpose, user expectations, versioning, and residual information require a dataset-specific assessment of remediation, as they generalize poorly across all ``public repositories.''